\documentclass[12pt,english,aps,manuscript]{article}
\usepackage[T1]{fontenc}
\usepackage[latin9]{inputenc}
\usepackage{geometry}
\usepackage{amsmath}
\usepackage{amssymb}
\usepackage{esint}
\usepackage[numbers]{natbib}
\usepackage{babel}
\begin{document}
\begin{center}
\textbf{\Large{}Comments on Entropy Calculations in Gravitational
Systems }{\Large\par}
\par\end{center}

\begin{center}
\vspace{0.3cm}
\par\end{center}

\begin{center}
{\large{}S. P. de Alwis$^{\dagger}$ }{\large\par}
\par\end{center}

\begin{center}
Physics Department, University of Colorado, \\
 Boulder, CO 80309 USA 
\par\end{center}

\begin{center}
\textbf{Abstract} 
\par\end{center}

\begin{center}
\vspace{0.3cm}
\par\end{center}

\smallskip{}
\vspace{0.3cm}
 We discuss the logic of, and some puzzles in, the various approaches
to thermodynamics of gravitational systems. In particular the blackhole,
deSitter (dS), black hole in dS (SdS) and in Anti-deSitter SAdS backgrounds
are considered. After reviewing the original calculations of Hawking
and Gibbons we discuss an alternative Hamiltonian method. This justifies
the lowest order Euclidean calculation but is free of the problems
associated with the latter when going to higher orders. To conclude
we address the sign issue in dS thermodynamics.

\vfill{}

$^{\dagger}$ dealwiss@colorado.edu 

\eject

\section{Summary}

We begin by summarizing the classic calculations of Gibbons and Hawking
\citep{Gibbons:1976ue} (GH1) and \citep{Gibbons:1977mu} (GH2).
\begin{itemize}
\item The original (Euclidean path integral) calculation in GH1 for the
black hole action, got (with $\hbar=c=G=1$), $I=\beta F=\beta r_{{\rm h}}/4=\beta M-\frac{\beta}{\beta_{{\rm h}}}A_{{\rm h}}/4$
leading to the identification $S=A_{{\rm h}}/4=4\pi M^{2}$ if we
identify the average energy in a canonical ensemble as $<E>=M$. Here
the subscript ${\rm h}$ denotes horizon and $\beta_{{\rm h}}=\frac{2\pi}{\kappa_{{\rm h}}}=4\pi r_{{\rm h}}$
with $\kappa_{{\rm h}}$ being the surface gravity at the horizon
(as measured at infinity!), and $r_{h}=2M$ is the horizon radius
. For general beta the Euclidean metric has a conical singularity
which is removed precisely at $\beta=\beta_{h}$ i.e. the Hawking
temperature. The original derivation of the temperature was however
based on using the behavior of quantum fields in a black hole (of
dS) space-time \citep{Hawking:1975vcx}\citep{Gibbons:1977mu} and
did not depend on Euclidean path integral methods.
\item However this result for the entropy forces us to conclude that the
canonical ensemble makes no sense - it is divergent since the density
of states rises as $\exp\left(8\pi E^{2}\right)$, assuming the entropy
of a black hole of energy (mass) $E$ is given by the usual formula
above, overwhelming the Boltzmann factor $\exp\left(-\beta E\right)$.
One can as advocated by York\citep{York:1972sj,Brown:1994su} insert
a wall at a finite distance from the horizon, that is coupled to a
heat bath at a fixed temperature to define equilibrium thermodynamics
- but that still does not get rid of the problem with the density
of states. However as pointed out in \citep{Brown:1992bq} the microcanonical
ensemble is well-defined. Furthermore in Schwarzchild Anti-deSitter
(SAdS) space the entropy grows only as $E^{2/3}$ so that the canonical
ensemble makes sense \citep{Hawking:1982dh}.
\item In GH1 the BH entropy comes entirely from the GHY boundary term (at
infinity) since the bulk term vanishes for a (Euclidean) vacuum solution.
In the dS case (in the static patch) on the other hand the entire
classical action comes from the bulk term since there is no boundary
at $r=0$.
\item For the extended (Kruskal-Szekeres) diagram for the black hole as
well as for dS space in global coordinates the entropy (at least to
leading order) is zero. This is consistent with identifying these
space times with quantum mechanical pure states \citep{Martinez:1994ub}. 
\item It is unclear how to define the thermodynamics of SdS space since
the two horizons are at different temperatures $\kappa_{{\rm b}}/2\pi\ne\kappa_{c}/2\pi$.
Gibbons and Hawking \citep{Gibbons:1977mu} advocate putting a perfectly
reflecting wall in-between the two horizons. But this is artificial
and surely not the same as the original system. The system itself
is thermodynamically unstable \citep{Bousso:1996au,Teitelboim:2001skl}. 
\end{itemize}

\section{Review of original calculations and comments thereof}

We will first quote some basic formulae\footnote{See for example the excellent account given by Eric Poisson \citep{Poisson:2009pwt}.}
for the gravitational action in both Lagrangian and Hamiltonian form.
Note that we work in units $\hbar=c=G=1$ and the mostly plus metric
convention, with $R_{\mu\nu\,\,\,\sigma}^{\,\,\,\,\,\lambda}V^{\sigma}\equiv\left[\nabla_{\mu},\nabla_{\nu}\right]V^{\lambda}$
and $R_{\nu\sigma}=R_{\,\,\,\,\nu\mu\sigma}^{\mu}$. The Einstein-Hilbert
action for gravity (without a cosmological constant) is
\begin{equation}
I_{H}\left[g_{\alpha\beta}\right]=\frac{1}{16\pi}\int_{{\cal V}}R\sqrt{-g}d^{4}x.\label{eq:SH}
\end{equation}
Here ${\cal V}$ is a space-time volume with a boundary which will
be denoted by $\partial{\cal V}$. Under a metric variation we have
\begin{equation}
16\pi\delta I_{H}=\int_{{\cal V}}\left(R_{\alpha\beta}-\frac{1}{2}g_{\alpha\beta}R\right)\delta g^{\alpha\beta}\sqrt{-g}d^{4}x-\int_{\partial{\cal V}}\epsilon h^{\alpha\beta}\delta g_{\alpha\beta,\mu}n^{\mu}|h|^{1/2}d^{3}y.\label{eq:dSH}
\end{equation}
The first term contains the Einstein tensor which is the LHS of the
Einstein field equation. In the boundary term the $y$ are coordinates
adapted to $\partial{\cal V}$, $n^{\mu}$ is the unit normal to it
with $n^{2}=\epsilon$, where $\epsilon=\pm1$ depending on whether
the boundary component is time-like or space-like, and $h$ is the
metric on $\partial{\cal V}$ with $h^{\alpha\beta}=g^{\alpha\beta}-\epsilon n^{\alpha}n^{\beta}$.
If we impose Neumann (N) boundary conditions on $\partial{\cal V}$
the second term vanishes and the variational principle will lead to
the field equations.

However if we wish to impose Dirichlet (D) boundary conditions on
$\partial{\cal V}$, we need to add a boundary term whose variation
(after fixing the metric on the boundary i.e. $\delta g_{\alpha\beta}|_{\partial{\cal V}}=0$
will cancel the boundary term in \eqref{eq:dSH}, thus giving the
desired equation of motion. This term \citep{Gibbons:1976ue} is
\begin{equation}
I_{B}\left[g_{\alpha\beta}\right]=\frac{1}{8\pi}\int_{{\cal \partial V}}\epsilon K|h|^{1/2}d^{3}y.\label{eq:SB}
\end{equation}
Here $K=\nabla_{\mu}n^{\mu}$ is the trace of the so-called second
fundamental form (i.e. the extrinsic curvature of $\partial{\cal V}$,
$K_{\mu\nu}=\nabla_{\mu}n_{\nu}$ as embedded in ${\cal V}$ where
$n_{\mu}$ is the unit normal to $\partial{\cal V}$). Thus the appropriate
action that should be used in the path integral is
\begin{equation}
I=I_{H}+I_{B}-I_{B_{0}},\label{eq:S}
\end{equation}
 where the last term is given by \eqref{eq:SB} with $K\rightarrow K_{0}$
i.e. the extrinsic curvature evaluated in flat space. This is required
in order to have zero action for flat space. Clearly it will not contribute
to the equation of motion. It is important to note that this implies
(according to the entropy calculation of Gibbons and Hawking) empty
Minkowski space (the vacuum) will be assigned zero entropy consistent
with standard thermodynamics in flat space where the ground state
is assigned zero entropy.

One might ask why one would need to impose D boundary conditions rather
than N. The reason is that (as explained for instance in \citep{Hawking:1979ig})
if one wishes to compute the quantum amplitude for transitions from
one metric $h_{1}$ on a surface (1) to another $h_{3}$ on another
surface (3), then in order that the sum over states with intermediate
metrics $h_{2}$ on a surface (2) gives the correct relation in the
classical limit, we need $I(g_{1}+g_{2})=I(g_{1})+I(g_{2})$, where
$g_{1}$ is the metric between the surfaces 1 and 2 and $g_{2}$ being
the metric between 2 and 3. This relation can only be satisfied with
D boundary conditions.

Given this action for gravity, namely eqn. \eqref{eq:S}, the thermodynamic
interpretation given in \citep{Gibbons:1976ue,Gibbons:1977mu} follows.

The discussion in \citep{Gibbons:1976ue} (GH1) actually begins by
evaluating the Lorentzian action \eqref{eq:S}. GH1 first evaluate
it for the Schwarzschild black hole with the metric 
\begin{equation}
ds^{2}=-\left(1-\frac{r_{s}}{r}\right)dt^{2}+\left(1-\frac{r_{s}}{r}\right)^{-1}+r^{2}d\Omega_{2}^{2},\label{eq:BHmetric}
\end{equation}
where the last term is the metric on a two sphere with radius $r$
and the Schwarzschild (horizon) radius $r_{s}$ may be identified
as $2M$ where $M$ is the mass of the black hole.

This has a curvature singularity at $r=0$ as well as a coordinate
singularity at $r=r_{s}$. In view of these the question is how does
one evaluate the bulk action $S_{H}$? GH1 prescription is to evaluate
the action in the so-called Euclidean section where $t\rightarrow-i\tau$
with $\tau$ taken to be real. The metric is then everywhere smooth
provided that $\tau$ is taken to be an angular coordinate with $\tau$
and $\tau+4\pi r_{s}$ identified. In this case the bulk term $S_{H}$
is zero since $R=0$ everywhere, and the classical action comes entirely
from the boundary term which now has the topology $S^{1}\times S^{2}$
and is compact.

To evaluate this we need the unit normal to the boundary $r={\rm constant}$
which is $n_{\mu}=\delta_{\mu}^{r}/\sqrt{1-\frac{r_{s}}{r}}$, $n^{\mu}=g^{\mu\nu}n_{\nu}=\sqrt{1-\frac{r_{s}}{r}}\delta_{r}^{\mu}$,
so we get $K=\nabla.n=\frac{1}{\sqrt{g}}\partial_{\mu}\left(\sqrt{g}n^{\mu}\right)=\frac{1}{r^{2}}\partial_{r}\left(r^{2}\sqrt{1-\frac{r_{s}}{r}}\right)$
and $K_{0}=\frac{1}{r^{2}}\partial_{r}\left(r^{2}\right)=\frac{2}{r}$.
Also we have $|h|^{1/2}=\sqrt{1-\frac{r_{s}}{r}}r^{2}\sin\theta$.
Hence (defining the boundary at $r=r_{b}\rightarrow\infty)$
\begin{align}
I_{{\rm bh}} & =\frac{1}{8\pi}\int_{0}^{-i\beta}dt\int_{S_{2}}d^{2}x|h|^{1/2}\left(K-K_{0}\right)|_{r=r_{{\rm b}}}=-i\frac{\beta}{2}\left(2r_{{\rm b}}-\frac{3}{2}r_{s}-2r_{{\rm b}}\sqrt{1-\frac{r_{s}}{r_{{\rm b}}}}\right)\nonumber \\
 & =i\frac{\beta}{2}\left(\frac{r_{s}}{2}+O\left(\frac{r_{s}^{2}}{r_{{\rm b}}}\right)\right)\overset{\beta=4\pi r_{s},\,r_{{\rm b}}\rightarrow\infty}{=}i\pi r_{s}^{2}\label{eq:bhaction}
\end{align}
Note that the the final result is $i$ times a quarter the area of
the horizon. 

GH1 also calculate the classical action for deSitter space. In this
case \eqref{eq:SH} needs to be replaced by 
\begin{equation}
I_{H}\left[g_{\alpha\beta}\right]=\frac{1}{16\pi}\int_{{\cal V}}\left(R-\frac{6}{r_{{\rm ds}}^{2}}\right)\sqrt{-g}d^{4}x,\label{eq:SHCC}
\end{equation}
where $r_{{\rm ds}}^{2}=3/\Lambda$ and $\Lambda$ is the (positive)
cosmological constant. Also the space-time has no boundary so there
is no boundary term and no singularity. From the Einstein equation
we have $R=12/r_{{\rm ds}}^{2}$. However if we substitute this into
the action we get infinity from the infinite space time volume. 

GH1 argues that one has to again consider the Euclidean section. First
let us consider the so-called static patch representation of (part
of) dS space given by a presentation like that of the black hole in
\eqref{eq:BHmetric}:
\begin{equation}
ds^{2}=-\left(1-\frac{r^{2}}{r_{{\rm ds}}^{2}}\right)dt^{2}+\left(1-\frac{r^{2}}{r_{{\rm ds}}^{2}}\right)^{-1}dr^{2}+r^{2}d\Omega_{2}^{2}.\label{eq:dSmetric}
\end{equation}
As with the black hole metric presentation in \eqref{eq:BHmetric},
the above is also singular at the horizon $r=r_{{\rm ds}}$ (where
the time-like Killing vector $\partial/\partial t$ becomes null).
Again following the prescription used in the BH case GH1 propose evaluating
the action in the Euclidean section with $t\rightarrow-i\tau$ (with
$\tau$ real). Also as in the previous case the relation between these
coordinates and the global coordinates implies that the Euclidean
variable $\tau$ is periodic with period $2\pi r_{{\rm ds}}$,. This
also avoids the conical singularity at $r=r_{{\rm ds}}$ in the Euclideanized
version of \eqref{eq:dSmetric}. The topology of the Euclidean section
is now an $S_{4}$ with radius $r_{{\rm ds}}$, so the action becomes,
\begin{align}
iI_{{\rm ds}} & \rightarrow-I_{{\rm ds}}^{{\rm Euclidean}}=-\frac{1}{16\pi}\int\sqrt{|g|}d^{4}x_{{\rm Euclidean}}\left(\frac{6}{r_{{\rm ds}}^{2}}-R\right)=\nonumber \\
 & =\frac{1}{16\pi}\int\sqrt{|g|}d^{4}x_{{\rm Euclidean}}\frac{6}{r_{{\rm ds}}^{2}}=\frac{1}{16\pi}\frac{8\pi^{2}r_{{\rm ds}}^{4}}{3}\frac{6}{r_{{\rm ds}}^{2}}=\pi r_{{\rm ds}}^{2}.\label{eq:dSaction}
\end{align}
 What do these calculations have to do with thermodynamics? This follows
according to GH1 from the representation of the partition function
for the canonical ensemble by a path integral, i.e.
\begin{equation}
Z\left(\beta\right)={\rm Tr}e^{-\beta H}=\int_{\phi\left(t\right)=\phi(t+i\beta)}\left[d\phi\right]e^{iI[\phi]}\label{eq:Z}
\end{equation}
Here $\phi$ stands for all fields in the system including the metric
field. One then evaluates the RHS in the saddle point approximation
by writing $\phi=\phi_{0}+\bar{\phi}$ in which case we have,
\begin{equation}
-\beta F\equiv\ln Z\simeq iI\left[\phi_{0}\right]+\ln\int_{\bar{\phi}\left(t\right)=\bar{\phi}(t+i\beta)}\left[d\bar{\phi}\right]e^{iI[\phi_{0}\bar{,\phi}]}+\ldots.\label{eq:lnZ}
\end{equation}
Here the second term is the one loop correction and the ellipses represent
higher loop terms.

Let us evaluate this for the two systems discussed above, i.e. the
Schwarzschild black hole and deSitter space. In both cases $\phi$
is just the metric field. For the black hole (implicitly assuming
that $\phi_{0}$ just represents a unique black hole of mass $M$)
and ignoring the one-loop correction we have,
\begin{equation}
\beta F_{0}=-iI_{{\rm bh}}[g_{0}]=\pi r_{s}^{2}=\beta<E>-S\label{eq:FES}
\end{equation}
In the second equality we used \eqref{eq:bhaction} and the last one
is the usual thermodynamic relation between free energy $F$ the average
energy of the ensemble $<E>$ and the entropy $S$. GH1 then proceeds
to identify $<E>$ with the mass $M$, and with $\beta=4\pi r_{s},\,M=r_{s}/2$
this gives the famous formula $S=\pi r_{s}^{2}=\frac{1}{4}A_{{\rm h}}$.
A similar argument for dS space gives using\eqref{eq:dSaction}
\begin{equation}
\beta F_{0}=I_{{\rm ds}}^{{\rm Euclidean}}=-\pi r_{{\rm ds}}^{2}=\beta<E>-S\label{eq:FES2}
\end{equation}
In this case GH1 identify $<E>$ with zero since space is closed (so
there is no boundary energy) giving again $S=\frac{1}{4}A_{{\rm h}}.$

At this point it is worthwhile noting that the two calculations apparently
give contradictory results for the entropy of empty Minkowski space.
The $M\rightarrow0$ limit of black hole space time gives Minkowski
space. Given that the calculation of GH1 gave $S_{{\rm bh}}=\pi r_{s}^{2}=\pi\left(2M\right)^{2}\rightarrow0$
in the limit, this implies that the entropy of Minkowski space is
zero. This is clearly consistent with the expectation that quantum
mechanically this geometry must represent a vacuum state. On the other
hand one might also argue that Minkowski space is obtained from deSitter
space by taking the limit $r_{{\rm ds}}\rightarrow\infty$ so that,
since GH1 gave $S_{{\rm ds}}=\pi r_{{\rm ds}}^{2}$, one has infinite
entropy in the limit. In other words one is apparently led to assign
infinite entropy to (empty) Minkowski space, which seems rather strange!
However It has been argued by many authors (perhaps the earliest were
\citep{Banks:2000fe} and \citep{fischler2000taking}), that the horizon
entropy of dS space is the maximum entropy that this space can hold.
With this interpretation then, in the limit of the horizon radius
going infinity, the ensuing infinite entropy of Minkowski space, should
also be interpreted as the maximum entropy that this space can contain,
which of course is reasonable. It is not therefore the entropy of
the Minkowski vacuum, which should indeed be zero.

If we look at the calculations in GH1 we see that the two entropies
arise from quite different terms in the (Euclidean) action. In the
black hole case the space-time volume integral (the Einstein-Hilbert
term) gives zero contribution since the space is Ricci flat. The entropy
comes from the boundary term. As discussed above, GH1 subtracts from
the boundary term a term integrated over the same surface but with
$\nabla.n$ evaluated in flat space. It is the cancellation between
these two for Minkowski space which results in zero entropy for this
space. On the other hand in the corresponding calculation for dS space
it is the volume integral which gives the entropy and there is no
GHY boundary term since there is no boundary. Now the integrand in
this term (see eqn. \eqref{eq:dSaction}) goes to zero like $\frac{1}{r_{{\rm ds}}^{2}}$
in the Minkowski limit but the volume integral gives a factor of $r_{{\rm ds}}^{4}$
so the action and hence the entropy is proportional to $r_{{\rm ds}}^{2}$
! However the infinite radius limit of dS space should not be identified
with Minkowski space. The topologies are different and hence the need
for a (spatial) boundary term in the latter case which is absent in
the former. Indeed if one worked in the strict Minkowski limit (even
though the volume is still infinite) the curvature is strictly zero,
so that the E-H term is exactly zero. 

\subsection{Canonical and Microcanonical Ensemble for Black Holes\label{subsec:Canonical-and-Microcanonical}}

The canonical ensemble refers to a collection of systems in thermal
equilibrium at some externally fixed temperature $T=\beta^{-1}$.
Instead of a single energy one has a (Boltzmann) distribution of energies.
However as is well known the canonical partition function for Schwarzchild
black holes is ill-defined even with Brown-York boundary conditions
\citep{Brown:1992bq}, (which impose a physical boundary at a finite
distance held at a fixed temperature).

The point is simply that the density of states $D(E)$ (the exponential
of the entropy) increases as $\exp\left(4\pi E^{2}\right)$ (if indeed
the entropy is given by the Bekenstein-Hawking formula). In the Brown-York
set up each black hole is enclosed in a spherical wall at some radius
$R>r_{H}=2GM$ held at a fixed temperature $T=\beta^{-1}$. Nevertheless
clearly the density of states overwhelms the Boltzmann factor and
hence the partition function is divergent. 

It is more appropriate to consider the micro-canonical ensemble (as
advocated in \citep{Brown:1992bq}). Typically this is an ensemble
with a small (or vanishingly small) energy range $E-dE,E+dE$. Entropy
is then defined to be $S(E)=\ln D(E)$ where $D(E)$ is the density
of states at energy $E$. Instead of \eqref{eq:lnZ} one should start
with the functional integral representation of $D(E)$. This is
\begin{equation}
D(E)={\rm Tr}\delta\left(E-\hat{H}\right)=\frac{1}{2\pi}{\rm Tr}\int dte^{i\left(E-\hat{H}\right)t}=\frac{1}{2\pi}\int_{-\infty}^{+\infty}dte^{iEt}{\cal N}\int_{\phi\left(0\right)=\phi(t)}\left[d\phi\right]\int\left[d\pi\right]e^{i\int_{0}^{t}dt'\left[\int_{\Sigma}\pi\dot{\phi}-H\right]}\label{eq:NE}
\end{equation}
Here we've used the path integral representation for the (trace of
the) unitary evolution operator, 
\[
{\rm Tr}e^{i\hat{H}t}={\cal N}\int_{\phi\left(0\right)=\phi(t)}\left[d\phi\right]\int\left[d\pi\right]e^{i\int_{0}^{t}dt'\int_{\Sigma}\pi\dot{\phi}-H]}.
\]

The normalization constant ${\cal N}$ is defined so that the trace
of the unit operator i.e. ${\rm Tr}\hat{I}={\cal N}\int_{\phi\left(0\right)=\phi(t=0)}\left[d\phi\right]\int\left[d\pi\right]$
is unity, which implies that we have regularized and renormalized
our definition of the trace such that, as we'll see below, the extended
black hole or deSitter space time (or in general any stationary asymptotically
flat or deSitter space time), has zero entropy corresponding to a
pure state. The entropy associated with a black hole, or that of de
Sitter in static coordinates must then be identified as entanglement
entropy. Actually it makes more sense to use the original Hamiltonian
representation for the path integral. In this case, we replace $\phi$
by the set $N,N_{a},h_{ab},\psi$, where we've used a $3+1$ split
of space time and defined fields on the space-like surface $\Sigma_{t}$
as discussed earlier. Then the expression for the density of states
becomes
\begin{equation}
D(E)=\frac{1}{2\pi}\int dte^{iEt}{\cal N}\int_{\phi\left(0\right)=\phi(t)}\left[dN\right]\left[dN_{a}\right]\left[dh_{ab}\right]\left[d\pi^{ab}\right]\left[d\psi\right]\left[d\pi_{\psi}\right]e^{i\int_{0}^{t}dt'\left[\int_{\Sigma}\pi^{ab}\dot{h}_{ab}+\pi_{\psi}\dot{\psi}-H\right]}.\label{eq:NE2}
\end{equation}
Here $N,N_{a}$ become Lagrange parameters imposing the secondary
constraints (the primary are $\pi_{N}=\pi_{N_{a}}=0$). The Hamiltonian
is given by\footnote{See for example Poisson \citep{Poisson:2009pwt} eqn. 4.61. $k$ is
defined in eqn.\eqref{eq:k} below. } with matter Hamiltonian ${\cal H}_{m}$ and momentum density $\nabla_{b}{\cal P}_{m}^{ab}$)
\begin{align}
H & =\frac{1}{16\pi}\int_{\Sigma_{t}}\left[N\left((K^{ab}K_{ab}-K^{2}-^{3}R)+{\cal H}_{{\rm m}}\right)-2N_{a}\nabla_{b}\left(K^{ab}-Kh^{ab}+{\cal P}_{{\rm m}}^{ab}\right)\right]\sqrt{h}d^{3}y+H_{\partial},\label{eq:H}\\
 & \equiv\int_{\Sigma_{t}}\left[N{\cal H}+N_{a}{\cal H}^{a}\right]+H_{\partial},\\
H_{\partial} & =\frac{1}{8\pi}\ointop_{S_{t}}\left[N\left(k-k_{0}\right)-N_{a}\left(K^{ab}-Kh^{ab}\right)r_{b}\right]\sqrt{\sigma}d^{2}\theta.\label{eq:Hboundary}
\end{align}
Here $S_{t}$ is the $S_{2}$ boundary of $\Sigma_{t}$. Doing the
functional integrals over the $N's$ gives
\begin{equation}
D(E)=\frac{1}{2\pi}\int dte^{iEt}{\cal N}\int_{\phi\left(0\right)=\phi(t)}\left[dh_{ab}\right]\left[d\pi^{ab}\right]\left[d\psi\right]\left[d\pi_{\psi}\right]\delta\left[{\cal H}\right]\delta\left[{\cal H}^{a}\right]e^{i\int_{0}^{t}dt'\left[\int_{\Sigma_{t'}}\left(\pi^{ab}\dot{h}_{ab}+\pi_{\psi}\dot{\psi}\right)-H_{\partial}\right]}\label{eq:NE2-1}
\end{equation}
If we assume that the path integral is saturated by a single stationary
configuration with boundary Hamiltonian $H_{\partial}$ (as assumed
in\citep{Gibbons:1976ue}) then the $t$ integral also yields a delta
function and we have
\begin{equation}
D(E)=\delta\left(E-H_{\partial}\right){\cal N}\int\left[dh_{ab}\right]\left[d\pi^{ab}\right]\left[d\psi\right]\left[d\pi_{\psi}\right]\delta\left[{\cal H}\right]\delta\left[{\cal H}^{a}\right]|_{{\rm classical}}=\delta\left(E-H_{\partial}\right)e^{S(E)}.\label{eq:NE3}
\end{equation}
so that $\int_{H_{\partial}-\Delta E}^{H_{\partial}+\Delta E}D(E)dE=e^{S(H_{\partial})}.$ 

Thus the log of the coefficient of the delta function gives the entropy
of this configuration. i.e. the general expression for the exponential
of the entropy for a stationary configuration is
\begin{equation}
e^{S(H_{\partial})}={\cal N}\int\left[dh_{ab}\right]\left[d\pi^{ab}\right]\left[d\psi\right]\left[d\pi_{\psi}\right]\delta\left[{\cal H}\right]\delta\left[{\cal H}^{a}\right]|_{H_{\partial}}.\label{eq:NE3-1}
\end{equation}
Note that this formula is exact for stationary fluctuations around
the original stationary background. In general of course we need to
keep the time derivative terms in the exponent of \eqref{eq:NE2-1}.
In the above we've assumed that there is no inner boundary so that
the calculation is appropriate for the extended (two sided) black
hole. If one ignored the delta functions in the measure (they are
effectively a quantum correction), then we have in this approximation
$S(M)=0$ which is consistent with the picture of a two sided black
hole being a pure state as observed in \citep{Martinez:1994ub}\footnote{In this reference the normalization factor is effectively set equal
to one. However it is clear from the above that it has been implicitly
defined such that the trace of the unit operator is one.}. Note that the fields in the functional integral here are defined
as functions of global coordinates so that $\Sigma_{t}$ in the integrand
of \eqref{eq:NE2-1} is taken over both causal wedges (eg. region
I and III of the Kruskal diagram in the black hole case). The same
would be true of both dS and SAdS spaces. In other words the extended
field space for all these configurations should have entropy zero.
It is unclear how this generalizes to spaces with more than one horizon
such as SdS Kerr black holes etc..

\subsubsection{Schwarzchild black hole}

If on the other hand we wish to compute the entropy of one sided black
hole (here identified as say region $I$ of the diagram for an eternal
black hole) we need to compute in Schwarzschild coordinates in which
case (since they diverge at the horizon) we need to insert a boundary
there. Let us recall briefly\footnote{For details see for example \citep{Poisson:2009pwt} section 4.2.}
how the first term of \eqref{eq:Hboundary} arose. 

The boundary $\partial{\cal V}$ of the region of integration in \eqref{eq:SH}
can be split up into two space-like boundaries $\Sigma_{t_{1}},\Sigma_{t_{2}}$
and a time-like boundary ${\cal B}$. In the 1+3 split (ADM) formulation
the action for pure gravity becomes (with $I_{m}=\int\sqrt{|g|}{\cal L}_{m}$
being the matter action - including the cosmological constant),\footnote{\label{fn:notation}$n$ is the time-like normal to $\Sigma_{t}$
and $r$ is the space-like normal to ${\cal B}$, $y,z$ are respectively
coordinates on $\Sigma_{t},{\cal B}$ with metrics $h_{ab},\gamma_{ij}$.
Also ${\cal K}=\gamma^{ij}{\cal K}_{ij}=\gamma^{ij}\left(e_{i}^{\alpha}e_{j}^{\beta}\nabla_{\beta}r_{\alpha}\right)$.
One also has the completeness relation for the metric $g^{\alpha\beta}=r^{\alpha}r^{\beta}+\gamma^{\alpha\beta}=r^{\alpha}r^{\beta}-n^{\alpha}n^{\beta}+\sigma^{AB}e_{A}^{\alpha}e_{B}^{\beta}$.
Also note that note that ${\cal B}$ is foliated by $S_{t}$.} 
\begin{align}
16\pi I\left[g_{\alpha\beta}\right] & =\int_{{\cal V}}R\sqrt{-g}d^{4}x+2\int_{{\cal \partial V}}\epsilon K|h|^{1/2}d^{3}y+I_{{\rm m}}=\int_{t_{1}}^{t_{2}}dt\int_{\Sigma_{t}}\left(^{3}R+K^{ab}K_{ab}-K^{2}+{\cal L}_{m}\right)N\sqrt{h}d^{3}y\nonumber \\
 & +2\int_{t_{1}}^{t_{2}}dt\ointop_{S_{t}}\left({\cal K}+\nabla_{\beta}r_{\alpha}n^{\alpha}n^{\beta}\right)N\sqrt{\sigma}.\label{eq:I-3+1}
\end{align}
Note that the first term in the second line comes from the time-like
part of the GH boundary term \eqref{eq:dSH}. Using the completeness
relation for the metric (see footnote \eqref{fn:notation}) we then
have 
\begin{equation}
{\cal K}+\nabla_{\beta}r_{\alpha}n^{\alpha}n^{\beta}=\sigma^{AB}k_{AB}=k.\label{eq:k}
\end{equation}
 This boundary term in the action then gives us the first term of
the boundary Hamiltonian in \eqref{eq:Hboundary}.

On the inner boundary (at the black hole horizon) however there is
no GH boundary term thus the corresponding contribution to the boundary
integral there is $\frac{1}{8\pi}\ointop_{{\cal H}_{t}}\left[\nabla_{\beta}r_{\alpha}n^{\alpha}n^{\beta}\right]N\sqrt{\sigma}d^{2}\theta$
(there is a minus sign from the fact that the outward normal to ${\cal H}_{t}$
is $-r_{\alpha}$ and another one from going from the action to the
Hamiltonian). This is evaluated by first moving slightly away from
the horizon (to the stretched horizon) where the relevant vectors
($\hat{n},\hat{r}$ are respectively still time-like and space-like).
We have $n^{\mu}=N^{-1}\frac{dx^{\mu}}{dt}$ (normalized to $n^{2}=-1$)
and $r^{\mu}=N\frac{dx^{\mu}}{dr}$ (normalized to $r^{2}=1$) and
$r.n=0$. Then \footnote{All but the last two steps below are the same as in footnote 13 of
\citep{Banihashemi:2022jys} except that it's in Euclidean form.} 
\begin{align*}
Nn^{\mu}n^{\nu}\nabla_{\mu}r_{\nu} & =-Nn^{.u}r^{\nu}\nabla_{\mu}n_{\nu}=-Nn^{.u}r^{\nu}\nabla_{\mu}\left(-N\nabla_{\nu}t\right)=N^{2}n^{\mu}r^{\nu}\nabla_{\mu}\nabla_{\nu}t=N^{2}n^{\mu}r^{\nu}\nabla_{\nu}\left(-N^{-1}n_{\mu}\right)\\
 & =-N^{2}n^{\mu}r^{\nu}\nabla_{\nu}\left(N^{-1}\right)n_{\mu}=-N^{2}n^{\mu}r^{\nu}\left(-N^{-2}\nabla_{\nu}N\right)n_{\mu}=-r^{\nu}\nabla_{\nu}N\rightarrow-N\nabla_{r}N|_{r\rightarrow r_{{\cal H}_{t}}}=-\kappa,
\end{align*}
where we used $r^{\nu}\nabla_{\nu}t=0$, $n^{2}=-1,\,n^{\mu}\nabla_{\nu}n_{\mu}=0$,
and $\kappa$ is the surface gravity at the relevant horizon\footnote{The surface gravity $\kappa$ is defined by the acceleration at the
horizon, $\xi^{\beta}\nabla_{\beta}\xi^{\alpha}=\kappa\xi^{\alpha}$
and can be shown to be constant at the horizon. For static space times
it can be shown that $\kappa=N\frac{dN}{dr}$ where $r$is the radial
coordinate and $N=N(r)$. }.
\begin{equation}
\Delta H_{\partial}=\frac{1}{8\pi}\ointop_{{\cal H}_{t}}\left[\nabla_{\beta}r_{\alpha}n^{\alpha}n^{\beta}\right]N\sqrt{\sigma}d^{2}\theta=-\frac{1}{8\pi}\kappa{\cal A_{{\cal H}}}_{_{t}}.\label{eq:kappaA}
\end{equation}

Thus what one actually gets from the Lorentzian calculation is (in
the classical approximation as above with the path integral saturated
by a single black hole of mass $M=\frac{1}{8\pi}\ointop_{S_{t}^{\infty}}\left(k-k_{0}\right)N\sqrt{\sigma}$)
\[
D\left(E\right)=\delta\left(E-\left(M-\frac{1}{8\pi}\kappa{\cal A}\right)\right).
\]
How are we to interpret this result? The formula tells us that only
states satisfying the relation $M=E+\frac{1}{8\pi}\kappa{\cal A}$
contribute. Alternatively if one computed the partition function in
this approximation\footnote{Note that here we just have one fixed energy - since we are focused
on just one black hole. So the issue of the divergence of the partion
function does not arise. In other words we are simply identifying
what the free energy is corresponding to the one black hole configuration.} we have
\begin{equation}
Z=e^{-\beta F}=\int dED(E)e^{-\beta E}=e^{-\beta\left(M-\frac{\kappa}{2\pi}\frac{1}{4}{\cal A}\right)}.\label{eq:Zbh1}
\end{equation}
IF we identify $\kappa/2\pi$ as the temperature and $M$ as the internal
energy $U$ then comparing this with the thermodynamic relation $F=U-TS$
we would obtain the relation $S={\cal A}/4$ . However this classical
approximation to the Lorentzian path integral cannot give an independent
derivation of the Hawking temperature. This should not be surprising
since if we restored Planck's constant Hawking's expression for the
temperature is $\kappa\hbar/2\pi$. In other words the temperature
is a quantum effect and was derived by Hawking by computing the Green's
functions of quantum fields in a black hole background. One should
not expect to derive it in the classical approximation. 

Of course once we have the temperature and the identification of the
internal energy $U=<E>$ as the black hole mass $M$ one could have
integrated the first law $dU=TdS$ to get the entropy up to an arbitrary
constant i.e. $S=\frac{1}{4}{\cal A}/\hbar+{\rm constant}$. The above
argument fixes the constant to be zero.

\subsubsection{AdS black hole (SAdS)}

This has the metric 
\begin{equation}
ds^{2}=-\left(1-\frac{2M}{r}+\frac{r^{2}}{b^{2}}\right)dt^{2}+\left(1-\frac{2M}{r}+\frac{r^{2}}{b^{2}}\right)^{-1}dr^{2}+r^{2}d\Omega_{2}^{2}\label{eq:SAdSmetric}
\end{equation}
Here $b\equiv\sqrt{3/|\Lambda|}$ is the AdS radius. For $M=0$ we
have (global) AdS while for $b\rightarrow\infty$ we have the Schwarzchild
black hole. There is only one horizon in this space time which can
be identified with the horizon of the black hole, i.e. the solution
of $\left(1-\frac{2M}{r}+\frac{r^{2}}{b^{2}}\right)=0$. Calling this
$r_{+}$ we have the relation 
\begin{equation}
2M=r_{+}\left(1+\frac{r_{+}^{2}}{b^{2}}\right).\label{eq:M-r+}
\end{equation}
The thermodynamics of this space time using Euclidean methods was
worked out in \citep{Hawking:1982dh}. The Hawking temperature calculation
gives\footnote{As is well-known the quick and easy way of getting the temperature
is to Euclideanize the metric and then require the absence of a conical
singularity at $r=r_{+}$ giving $\beta=\frac{4\pi}{\left(dN^{2}/dr\right)_{r=r_{+}}}$.
However this is equivalent to the original Lorentzian calculation
of Hawking in terms of QFT in a curved background so is independent
of Euclidean methods.}
\begin{equation}
\beta=\frac{4\pi r_{+}b^{2}}{b^{2}+3r_{+}^{2}}.\label{eq:betaSAdS}
\end{equation}
Note that $\beta<\beta_{0}\equiv\frac{2\pi b}{\sqrt{3}}$ (i.e. $T>T_{0}\frac{\sqrt{3}}{2\pi b}$).
Also $T\rightarrow\infty$ for $r_{+}=0$ and $r_{+}\rightarrow\infty$
and the temperature is minimum at $r_{+}=r_{0}\equiv\frac{b}{\sqrt{3}}$.
Unlike in flat space in AdS a black hole can be in thermodynamic equilibrium
with radiation without artificial reflecting walls. Note however that
for a given temperature there are two possible solutions for the horizon
radius $r_{+}.$ The larger value corresponds to a black hole with
positive specific heat (unlike its flat space analog) and hence is
stable, while the smaller value has negative specific heat and is
unstable to decay into a bath of radiation. This is easily seen from
the graph of temperature vs. radius. So for $r_{+}>r_{0}$ we have
a stable black hole whilst for $r_{+}<r_{0}$ the black hole is unstable
and will decay into radiation. This is the so-called Hawking-Page
phase transition. 

As for the entropy calculation, in our Hamiltonian framework it proceeds
as in the flat space case except that for $k_{0}$ in \eqref{eq:Hboundary}
one now has $k$ evaluated in the background AdS space rather than
in flat space and the horizon radius and inverse temperatures are
given by \eqref{eq:M-r+} and \eqref{eq:betaSAdS}. Thus the formula
for the free energy is again given by \eqref{eq:Zbh1} so that the
entropy (after identifying $\beta=\frac{2\pi}{\kappa}$) is again
\begin{equation}
S=\frac{1}{4}{\cal A}=\pi r_{+}^{2},\label{eq:EntropySAdS}
\end{equation}
with the horizon radius given in terms of the black hole mass by \eqref{eq:M-r+}.
The latter formula also tells us that for large $r_{+},M$, the entropy
goes as $M^{2/3}$. So unlike in the flat space case, the density
of states does not swamp the Boltzmann term $-\beta M$ and the canonical
ensemble is well-defined.

\subsubsection{deSitter space (dS)}

Now consider dS. Again as in the black hole case if we identify the
spatial slice $\Sigma_{t}$ as extending over the whole spatial region
of global dS we would get zero entropy (at least classically as in
the black hole case). On the other hand if we take $\Sigma_{t}$ as
extending only over the static patch (i.e. from the dS horizon to
$r=0$), we get (since the only boundary is at the dS horizon ($r=0$
is the origin of coordinates!), 
\[
D\left(E\right)=\delta\left(E+\frac{1}{8\pi}\kappa_{c}{\cal A}_{c}\right).
\]

Again to interpret this we need to go back to the partition function
so,
\begin{equation}
Z=e^{-\beta F}=\int dED(E)e^{-\beta E}=e^{-\beta\left(-\frac{\kappa_{c}}{2\pi}\frac{1}{4}{\cal A}_{c}\right)}.\label{eq:ZdS1}
\end{equation}
 As before identifying $\kappa_{c}/2\pi$ as the temperature of the
horizon \citep{Gibbons:1977mu} and comparing with the formula $F=U-TS$
but now with the internal energy\footnote{There is no boundary energy associated with dS, so Gibbons and Hawking
take it to be zero \citep{Gibbons:1976ue}} $U=0$ we get the Gibbons Hawking result $S=\frac{1}{4}{\cal A}_{c}$.
In this case there is no direct way of comparing with the thermodynamic
identity $dU=TdS$ since the energy is fixed at zero (actually it
is not defined). The Gibbons-Hawking argument proceeds by adding external
energy to the system and identifying it as that which goes into the
thermodynamic identity $dU$. We will discuss the implications of
this below after rederiving the geometric version of the thermodynamic
identity. 

In computing the surface gravity at the dS horizon from the formula
$\kappa=\frac{1}{2}\nabla_{r}N^{2}$, it should be noted that this
gives the right sign for a black hole where $N^{2}=1-r_{s}/r$, $\kappa=1/2r_{s}$,
whereas for dS space, where $N^{2}=1-r^{2}/r_{{\rm dS}}^{2}$, since
the direction of increasing $r$ is the opposite of the black hole
case, $\kappa=-\frac{1}{2}\nabla_{r}N^{2}$, giving $\kappa=1/r_{{\rm dS}}$. 

\subsubsection{deSitter black hole (SdS)}

Now let us consider Schwarzschild deSitter (SdS) space. The metric
is 
\begin{equation}
ds^{2}=-\left(1-\frac{2M}{r}-\frac{\Lambda}{3}r^{2}\right)dt^{2}+\left(1-\frac{2M}{r}-\frac{\Lambda}{3}r^{2}\right)^{-1}dr^{2}+r^{2}d\Omega_{2}^{2}.\label{eq:SdS}
\end{equation}
Now there is a black hole horizon at $r=r_{{\rm b}}$ as well as a
cosmological horizon at $r=r_{{\rm c}}$ (when $\Lambda>0,\,9\Lambda M^{2}<1$)\footnote{If the last inequality is not satisfied the two real positive roots
become complex and there are no horizons.}, whose positions are given by the two positive roots of $r-2M-\frac{\Lambda}{3}r^{3}=0$.
The three roots are such that $r_{-}<0<2M<r_{{\rm b}}<3M<r_{{\rm c}}$
and $N^{2}=-\frac{1}{r}\frac{\Lambda}{3}\left(r-r_{-}\right)(r-r_{{\rm b}})(r-r_{{\rm c}})$.
The maximum value of the black hole mass is $M=\Lambda^{-1/2}/3$
where the two horizons coincide - this is the so-called Narai space.
From the above formula we then have, with the upper(lower) sign for
blackhole(cosmological) horizons, because of the different directions
in which the horizon normals point with respect to the direction of
increasing $r$.
\begin{equation}
T_{{\rm b,c}}=\frac{1}{2\pi}\kappa_{{\rm b,c}}=\pm\frac{1}{2\pi}\frac{1}{2}\left(\frac{2M}{r_{{\rm b,c}}^{2}}-\frac{\Lambda}{3}2r_{{\rm b,c}}\right).\label{eq:Tbc}
\end{equation}
There are now two temperatures one associated with the black hole
horizon and the other with the cosmological horizon. The thermodynamics
of such are situation is unclear. In fact the system is not in thermal
equilibrium and is unstable \citep{Bousso:1996au}\citep{Teitelboim:2001skl}.

\subsection{Alternative Lorentzian arguments}

\subsubsection{Lorentzian conical singularity argument}

Marolf \citep{Marolf:2022ybi} has given a Lorentzian argument for
the (Schwarzchild) black hole entropy formula. Let us first note that
in the Hamiltonian framework, after imposing the constraints it is
clear that there is no singularity in the integrand for the action,
since (for a static or stationary metric) the Hamiltonian version
of the action consists of the boundary terms identified in the previous
subsection.

Alternatively let us take the Lagrangian in 3+1 split form given in
eqn. \eqref{eq:I-3+1}. On a classical solution one must have ${\cal H}K^{ab}K_{ab}-K^{2}-^{3}R)=0$.
Thus we have
\begin{align*}
16\pi I_{{\rm cl}} & =2\int_{t_{1}}^{t_{2}}dt\int_{\Sigma_{t}}\left(K^{ab}K_{ab}-K^{2}\right)N\sqrt{h}d^{3}y\\
 & -2\int_{t_{1}}^{t_{2}}dt\ointop_{S_{t}}\left({\cal K}+\nabla_{\beta}r_{\alpha}n^{\alpha}n^{\beta}\right)N\sqrt{\sigma}.
\end{align*}
Again it is clear that the Lorentzian action is well defined on a
solution. 

Nevertheless it has been argued in \citep{Marolf:2022ybi} that the
functional integral is better approximated by slices that are not
solutions but contain conical singularities. By contour deformation
into the complex plane one can still pick up a saddle point and it
turns out to be the Euclidean black hole solution. It would be interesting
to see whether this procedure can be applied to dS, SdS and SAdS spaces
as well.

\subsubsection{Algebraic QFT argument}

In a recent work Chandrasekaran Pennington and Witten\citep{Chandrasekaran:2022eqq}
have discussed black hole thermodynamics from an Algebraic QFT perspective.
Also these authors and Longo \citep{Chandrasekaran:2022cip}, have
discussed the thermodynamics of dS spaces from the same point of view.
Apparently the Von Neumann algebras in the two cases are different,
which seems rather curious. The arguments are rather abstract and
it is not clear to us how it relates to our Hamiltonian argument or
to the original Euclidean calculations. Nevertheless it is important
to understand the connection. We leave further discussion of this
to future work.

\subsection{Euclidean calculation}

At this point it behoves us to revisit the Euclidean argument. Here
one argues that the partition function can be represented as a functional
integral of a Euclidean action. So we write
\begin{align*}
Z & ={\rm Tr}e^{-\beta H}={\cal N}\int_{\phi\left(\tau=0\right)=\phi(\tau=\beta)}\left[d\phi\right]\int\left[d\pi\right]e^{\int_{0}^{\beta}d\tau'\left[\int_{\Sigma}\pi\dot{\phi}-H\right]}\\
 & ={\cal N}\int_{\phi\left(\tau=0\right)=\phi(\tau=\beta)}\left[d\phi\right]\int\left[d\pi\right]e^{\int_{0}^{\beta}d\tau'\left[\int_{\Sigma}\left(\pi\dot{\phi}-N{\cal H}-N^{a}{\cal P}_{a}\right)-H_{\partial}\right]}
\end{align*}
Assuming again that the path integral is well-approximated by a single
stationary classical solution (which must necessarily satisfy ${\cal H={\cal P}}_{a}=0$),
we have 
\begin{equation}
Z=={\rm Tr}e^{-\beta H}\simeq{\cal N}\int_{\phi\left(\tau=0\right)=\phi(\tau=\beta)}\left[d\phi\right]\int\left[d\pi\right]e^{\int_{0}^{\beta}d\tau'\left[-H_{\partial}\right]}\label{eq:Eucl}
\end{equation}
The Euclidean metric only describes region I of the black hole or
just the static patch of deSitter space. The horizon, as is well-known,
has a conical singularity which needs to be resolved. In addition
to the boundary Hamiltonian coming from infinity , we have also the
term corresponding to \eqref{eq:kappaA}. This then gives for the
exponent in \eqref{eq:Eucl} $-\frac{1}{8\pi}\int_{0}^{\beta}\kappa{\cal A_{{\cal H}}}d\tau_{_{t}}$.
Putting $\kappa d\tau=d\theta$ the integral becomes $-\frac{1}{8\pi}\int_{0}^{\kappa\beta}{\cal A_{{\cal H}}}d\theta_{_{t}}$
and resolving the conical singularity amounts to making the range
of $\theta$ cover the full circle of polar coordinates so that $\kappa\beta=2\pi$
implying that\footnote{As mentioned before the temperature calculation necessarily involves
quantum mechanics. This is seen here from the fact that $\tau$ has
dimensions of a length (with $c=1$ but $\hbar\ne1$), whereas $\beta$
which is defined as an inverse temperature in the definition of $Z$
has dimensions of inverse energy and $\kappa$ has inverse length
dimensions. So restoring the factor of $\hbar^{-1}$ and $G$ in the
functional integral we have $\hbar\beta\kappa=2\pi$ giving $T=\kappa\hbar/2\pi$
and $S={\cal A}/4G\hbar$. } $T=\kappa/2\pi$ and the integral becomes $-\frac{1}{4}A_{{\cal H}}$
which from \eqref{eq:Eucl} may be identified with the entropy associated
with the corresponding horizon. 

All this is well-known of course - all we've done in the previous
subsection, is to reformulate the calculation in Hamiltonian framework.
However let us now consider the SdS case. There are two horizons with
two different temperatures\footnote{except in the degenerate Narai case where $r_{{\rm b}}=r_{{\rm c}}=\Lambda^{-1/2}$
and $T_{{\rm b}}=T_{{\rm c}}=0$} $T_{{\rm b}}=\kappa_{{\rm b}}/2\pi$ and $T_{{\rm c}}=\kappa_{{\rm c}}/2\pi$.
In this case there is no way to resolve both conical singularities.
This means that this Euclidean configuration will not satisfy the
Euclidean equation of motion, breaking down at one or other conical
sigularity depending on which one was resolved\citep{Teitelboim:2001skl}.
Indeed it is hard to envisage such a system as being in thermal equilibrium
unless as advocated by Gibbons and Hawking \citep{Gibbons:1977mu}
one inserts a perfectly reflecting wall between the two horizons.

The above calculation is completely equivalent to that using the Gauss-Bonnet
(GB) theorem\footnote{The Gauss Bonnet theorem is valid for compact two dimensional manifolds.
The argument of the above reference relies on the fact that in the
near horizon region the space time factorizes into a two dimensional
cone times an $S_{2}$.} presented in \citep{Banados:1993qp}. That argument although specifically
aimed at black holes can clearly be used in the dS case as well. However
it cannot be applied to an SdS space. As we've discussed above once
one has introduced the range of the Euclidean time variable to be
$\beta$ then the corresponding angular range is $\beta\kappa$. In
the application of the GB theorem one is using it for the complete
disc. But in the SdS case one can complete the disc by taking $\beta\kappa=2\pi$
at one horizon or the other but clearly not both, since there is only
parameter $\beta$ defining the ensemble\footnote{Here we disagree with \citep{Gregory:2013hja}. This paper claims
to derive the (admittedly compelling) formula $S_{{\rm SdS}}=\frac{1}{4}\left({\cal A}_{{\rm b}}+{\cal A}_{{\rm c}}\right)$
by using (in effect) the GB theorem. However as argued above one cannot
use it simultaneously at both horizon. In \citep{Spradlin:2001pw}
the entropy formula for SdS in 3 dimensions is derived. But this space
has only one horizon and hence only one temperature, so the above
problem does not arise.}.

\subsubsection{The problem with Euclidean arguments}

While the Euclidean calculations in GH1 seem to give physically meaningful
results as long as we just compute the leading (classical) contribution),
going beyond this with the functional integral in the Euclidean section
is problematic. This is due to the well known conformal factor problem.
Computing quantum corrections to the leading order expressions for
the entropy will inevitably involve confronting the wrong sign of
the kinetic terms for the conformal fluctuations. This is of course
well-known and indeed was pointed out in the original papers by Hawking
and collaborators (see for instance Hawking's essay in \citep{Hawking:1979ig}).
Nevertheless at the leading order the Euclidean argument gives the
correct result - in so far as it agrees with the thermodynamic identity
$dE=TdS$ when the calculation of the temperature in Hawking's original
paper which is the Lorentzian calculation is substituted for $T$.
Of course this determines $S$ only up to a constant. 

In our discussion (subsection \eqref{subsec:Canonical-and-Microcanonical})
and in particular eqn. \eqref{eq:NE2} we gave a Lorentzian formula
which in the leading order justifies the Euclidean calculation - including
the constant. Furthermore in principle this formula is well-defined
(at least perturbatively and subject to the treatment quantum gravity
as an effective theory below the cutoff (at Planck or string scale
or whatever theory UV completes Einstein's theory). In other words
the formula avoids the problems of the Euclidean path integral precisely
because of the delta functions in the measure which impose the constraints
in the quantum theory. 

\section{Smarr relations and the first law of Gravitational Thermodynamics
- identifying cosmological entropy}

In this section we first review both the Smarr relation and the related
first law of gravitational thermodynamics \citep{Bardeen:1973gs}
with the aim of clarifying a sign issue in its application to dS spacetime.

The Smarr relation for (Kerr) black holes is,

\begin{equation}
M-2\Omega_{H}J_{H}-\frac{1}{4\pi}\kappa_{H}{\cal A}_{H}=\frac{1}{4\pi}\int_{\Sigma}R_{\,\,\,\nu}^{\mu}t^{\nu}d\Sigma_{\mu}\label{eq:Sm1}
\end{equation}
Here $M$ is the total mass defined as an integral over the bounding
surface at infinity of the black hole space-time (taken to be asymptotically
flat) and $\Sigma$ extends from the event horizon to the sphere at
infinity. $\Omega_{H},\kappa_{H}$ are the angular velocity and the
surface gravity of the black hole horizon $H$. The corresponding
first law of black thermodynamics is 
\begin{align}
\delta M-\Omega_{H}\delta J_{H}-\frac{\kappa_{H}}{2\pi}\delta\left(\frac{1}{4}{\cal A}_{H}\right) & =\frac{1}{8\pi}\delta\int G_{\,\,\,\nu}^{\,u}t^{\nu}d\Sigma_{\mu}-\frac{1}{16\pi}\int G^{\mu\nu}\delta g_{\mu\nu}t^{\lambda}d\Sigma_{\lambda}\nonumber \\
 & =\delta\int T{}_{\,\,\,\nu}^{\,u}t^{\nu}d\Sigma_{\mu}-\frac{1}{2}\int T^{\mu\nu}\delta g_{\mu\nu}t^{\lambda}d\Sigma_{\lambda}.\label{eq:BHlaw}
\end{align}
Here $t^{\mu}$ is the time like Killing vector of the space-time
and the integrals are over the space-like surface between the black
hole horizon and the bounding surface at infinity, with $d\Sigma_{\mu}=n_{\mu}d\Sigma$
being the oriented surface element. $T_{\mu\nu}$ is the energy momentum
tensor (including a cosmological constant term\footnote{So for a pure cosmological constant term without dynamical matter
$T_{\mu\nu}=-\frac{1}{8\pi}\Lambda g_{\mu\nu}$.} or a possibly non-zero minimum of a scalar potential, of an external
(to the black hole horizon) matter distribution. Consider now SdS
space. Here the surface $\Sigma$ may be taken to be stretched between
the black hole horizon and the cosmological horizon. Denoting the
quantities evaluated at the latter by the subscript $c$ the above
two relations take the form\footnote{In the original derivation \citep{Bardeen:1973gs} the second term
in eqn. \eqref{eq:SdSlaw} was not given but was included in a subsequent
essay by B. Carter in \citep{Hawking:1979ig}.}
\begin{equation}
-\frac{1}{4\pi}\kappa_{c}{\cal A}_{c}-2\Omega_{H}J_{H}-\frac{1}{4\pi}\kappa_{H}{\cal A}_{H}=\frac{1}{4\pi}\int R_{\,\,\,\nu}^{\mu}t^{\nu}d\Sigma_{\mu}\label{eq:Sm2}
\end{equation}
and 
\begin{align}
-\frac{\kappa_{c}}{2\pi}\delta\left(\frac{1}{4}{\cal A}_{c}\right)-\Omega_{H}\delta J_{H}-\frac{\kappa_{H}}{2\pi}\delta\left(\frac{1}{4}{\cal A}_{H}\right) & =\frac{1}{8\pi}\delta\int G_{\,\,\,\nu}^{\,u}t^{\nu}d\Sigma_{\mu}-\frac{1}{16\pi}\int G^{\mu\nu}\delta g_{\mu\nu}t^{\lambda}d\Sigma_{\lambda}\nonumber \\
 & =\delta\int T{}_{\,\,\,\nu}^{\,u}t^{\nu}d\Sigma_{\mu}-\frac{1}{2}\int T^{\mu\nu}\delta g_{\mu\nu}t^{\lambda}d\Sigma_{\lambda}.\label{eq:SdSlaw}
\end{align}
Here $G_{\mu\nu}=R_{\mu\nu}-\frac{1}{2}R$ and we've used the Einstein
equation in the last line. The terms involving the angular velocity
correspond to chemical potential terms (such as $\mu dN$) in the
thermodynamic identity. Let us ignore these (or consider non-rotating
black holes). As discussed above, Hawking's QFT calculation established
that a black hole appears to an external observer at infinity as a
black body at a temperature $T_{H}=\kappa_{H}\hbar/2\pi$. A similar
argument by Gibbons and Hawking established that to a observer at
(say) the origin of static patch coordinates in dS would associate
a temperature $\kappa_{c}\hbar/2\pi$ to the cosmological horizon.
From \eqref{eq:BHlaw} after identifying $M$ as the total energy
of the system and setting for the moment the external matter-energy
to zero, we would have in the black hole case $\delta M=T_{H}\delta\left(\frac{1}{4\hbar}{\cal A}_{H}\right)$,
giving the well-known identification of the entropy of a black hole
to be $S=\frac{1}{4\hbar}{\cal A}_{H}$, after fixing the integration
constant to be zero. 

On the other hand the corresponding formula for the dS or SdS case
is \eqref{eq:SdSlaw}. Taking pure dS with some external matter the
formula becomes 
\begin{equation}
-\frac{\kappa_{c}}{2\pi}\delta\left(\frac{1}{4}{\cal A}_{c}\right)=\delta\int T{}_{\,\,\,\nu}^{\,u}t^{\nu}d\Sigma_{\mu}-\frac{1}{2}\int T^{\mu\nu}\delta g_{\mu\nu}t^{\lambda}d\Sigma_{\lambda}.\label{eq:dSlaw}
\end{equation}

Let us take $T^{\mu\nu}=-\Lambda g^{\mu\nu}$, so that initially we
have empty dS space. The second term is then $\Lambda\int g^{\mu\nu}\delta g_{\mu\nu}t^{\lambda}d\Sigma_{\lambda}=-\Lambda\int\frac{\delta(N\sqrt{h}}{N}d^{3}y$
which even for $\Lambda>0$, as is the case here, may be positive
or negative. However if the result of adding matter is the creation
of a small black hole then this additional term is zero. To see this
consider the SdS metric Taking $\delta g_{\mu\nu}$ to be the result
of changing $M\rightarrow M+\delta M$ (and then setting $M=0$) we
get
\[
g^{\mu\nu}\delta g_{\mu\nu}=0
\]
 (true even for $M\ne0$) since $\sqrt{g}$ is independent of $M$.
In this case from \eqref{eq:dSlaw} we have, after identifying $\frac{\kappa_{c}}{2\pi}=T_{c}$
as the temperature of the dS horizon, and $\delta\int T{}_{\,\,\,\nu}^{\,u}t^{\nu}d\Sigma_{\mu}=\delta E$
as the variation of Killing energy, and then comparing with the thermodynamic
identity $\delta E=T\delta S$, the variation of the entropy associated
with the horizon to be $\delta S_{c}=-\delta\left(\frac{1}{4}{\cal A}_{c}\right)$!

This is a long standing puzzle and various resolutions have been proposed.
A summary of these proposals has been given in \citep{Banihashemi:2022htw}.
The original paper \citep{Gibbons:1977mu} had one explanation for
it. This was clarified further in \citep{Spradlin:2001pw}\footnote{In \citep{Spradlin:2001pw} the entropy and the thermodynamic relation
for $dS_{3}$ was investigated. Following arguments of Gibbons and
Hawking the authors considered $SdS_{3}$ which only has a horizon
at $r_{{\rm b}}=\sqrt{1-8GM}$, and so correspondingly only one temperature
$T=\frac{1}{2\pi}\sqrt{1-8GM}$. They then integrate $dM=TdS$ and
fixing the integration constant so that the maximal mass black hole
(with $M=1/8G$) has zero entropy, which gives them $S=-{\cal A}_{{\rm b}}/4G$!
However as in the above discussion they argue that the correct interpretaion
of the thermodynamic identity is $d(-M)=TdS$ which then gives positive
entropy. See the discussion around figure 8 of that paper.}. The argument is based on the fact that in the space-like slice of
the global dS space is closed so that the total energy (which is a
boundary term because of the constraints) is zero. Suppose that the
observer is at the south pole of the $S_{3}$. Any added energy there
must be balanced by negative energy in the causal patch of the north
pole. The thermodynamic relation must reflect what from the observer's
point of view is happening beyond the horizon of the observer, i.e.
in the north pole patch. Thus the observer would conclude that in
that patch the added energy is negative and hence the relation\eqref{eq:dSlaw}
(after dropping the last term) should be read as
\begin{equation}
\frac{\kappa_{c}}{2\pi}\delta\left(\frac{1}{4}{\cal A}_{c}\right)=TdS=-\delta\int T{}_{\,\,\,\nu}^{\,u}t^{\nu}d\Sigma_{\mu}=\delta E\label{eq:dS1stlaw}
\end{equation}
 where $\delta E$ now refers to the observer's interpretation of
what goes on behind the horizon.

This argument has been criticized in \citep{Banihashemi:2022htw}
on the grounds that the time-like Killing vector in the north pole
patch is pointing downwards - meaning that in that patch the Killing
energy (which is the negative of the Killing energy in the south pole
patch) is actually positive. However this is not relevant to the thermodynamics
of the cosmological horizon of the south pole observer who must interpret
what is going on behind the horizon. In other words from the south
pole it appears as if the observed system, i.e. the one beyond the
south pole observer's horizon, has acquired negative energy and the
relevant relation is \eqref{eq:dS1stlaw}. Obviously observers on
the North pole will infer a similar relation for the thermodynamics
of what is beyond their horizon. There is of course an inevitable
subjectivity in these interpretations since the horizon itself is
observer dependent. The paper \citep{Banihashemi:2022htw} also argues
that if, as is expected, the state of global dS is a pure state (as
also argued in the comments after eqn. \eqref{eq:NE2-1} in this note),
then the entropy inside the horizon must equal the entropy outside
so both must decrease. However this is in contradiction with the argument
in \citep{Spradlin:2001pw} since the energy inside is obviously increasing.
But here again the issue is what does the entropy inside mean? What
\citep{Banihashemi:2022htw} refers to as entropy inside is actually
the horizon entropy as seen by the observer at the north pole. As
argued above this also decreases since there is complete symmetry
between the two observers. In other words once one interprets entropy
as representing a measure of the information inaccessible to each
observer, then there is no conflict\footnote{The work of \citep{Banihashemi:2022htw} advocates an alternate interpretation.
It is based on a somewhat different interpretaion of horizon entropy
than what is advocated in \citep{Spradlin:2001pw} and the present
work.}.

Now if we use \eqref{eq:dS1stlaw} to solve for the entropy we get
\[
S=\frac{1}{4}{\cal A}_{c}+{\rm constant}.
\]
How are we to fix this constant? In the black hole case the constant
was fixed by arguing that when the horizon area goes to zero (i.e.
the black hole mass is zero) we should get flat space, which should
have zero entropy, thus fixing the constant to be zero. In the dS
case however the horizon going to zero is the infinite cosmological
constant\footnote{${\cal A}_{c}=4\pi r_{{\rm ds}}^{2},$ where $r_{s}=\sqrt{3/\Lambda}$
so flat space is obtained in the limit $r_{s}\rightarrow\infty$.} case! This seems bizarre to say the least. On the other hand if we
had set the constant to be zero, as is usually done then as discussed
earlier, we would get the entropy of empty flat space (i.e. when $r_{{\rm ds}}\rightarrow\infty$)
to be infinite in sharp contradiction to the argument in the black
hole case where the entropy of flat space was set to zero! But as
we saw above this entropy is the maximum possible in dS space so its
Minkowski limit is the maximum entropy that that space can hold which
is plausibly infinite. However the point is that there is no plausible
argument for fixing this constant unlike in the black hole case. Our
Hamiltonian perspective however did fix both constants.

\section{Conclusion}

In these notes we've reviewed the entropy calculations for static
space-times with horizons, based on Euclidean arguments, and presented
alternative Lorentzian arguments. At leading order these serve to
justify the leading order Euclidean calculations. However unlike the
Euclidean calculations which suffer from the well-known problem of
the wrong sign kinetic term for conformal fluctuations, the Lozentzian
method is well-defined modulo the usual problems of perturbative quantum
field theories with gravity. In particular the formulae for the entropy
\eqref{eq:NE2-1} to \eqref{eq:NE3-1} are free of the conformal factor
issue since the constraints eliminate integration over these unphysical
modes. Approximating the relevant path integrals with a single black
hole etc. leads to the familiar expressions derived by Euclidean methods.

It should be pointed out that we do not have an independent derivation
of the Hawking temperature - for that we need to appeal to Hawking's
original Lorentzian calculations. Of course once one has the Hawking
temperature formulae one may use the thermodynamic identity to evaluate
changes in the entropy. Our calculation shows how the constant may
be fixed. This is particularly relevant in the dS case since (unlike
in the black hole case) the constant cannot be fixed by taking the
flat space limit. We also discussed the AdS black hole case and noted
some puzzles associated with SdS space.

Finally we revisited the geometric version of the first law and argued
in favor of the analysis of \citep{Spradlin:2001pw} (a modified version
of that in \citep{Gibbons:1977mu}) for interpreting the sign of this
relation in the dS case.

\section{Acknowledgements}

I wish to thank Sebastian Cespedes, Francesco Muia, and Fernando Quevedo
for discussions. I also wish to thank Don Marolf for clarifying the
arguments of \citep{Marolf:2022ybi} and to Ted Jacobson for email
correspondence on \citep{Banihashemi:2022htw}.

\bibliographystyle{apsrev}
\nocite{*}
\bibliography{myrefs}

\begin{thebibliography}{41}
\expandafter\ifx\csname natexlab\endcsname\relax\def\natexlab#1{#1}\fi
\expandafter\ifx\csname bibnamefont\endcsname\relax
  \def\bibnamefont#1{#1}\fi
\expandafter\ifx\csname bibfnamefont\endcsname\relax
  \def\bibfnamefont#1{#1}\fi
\expandafter\ifx\csname citenamefont\endcsname\relax
  \def\citenamefont#1{#1}\fi
\expandafter\ifx\csname url\endcsname\relax
  \def\url#1{\texttt{#1}}\fi
\expandafter\ifx\csname urlprefix\endcsname\relax\def\urlprefix{URL }\fi
\providecommand{\bibinfo}[2]{#2}
\providecommand{\eprint}[2][]{\url{#2}}

\bibitem[{\citenamefont{Gibbons and
  Hawking}(1977{\natexlab{a}})}]{Gibbons:1976ue}
\bibinfo{author}{\bibfnamefont{G.~W.} \bibnamefont{Gibbons}} \bibnamefont{and}
  \bibinfo{author}{\bibfnamefont{S.~W.} \bibnamefont{Hawking}},
  \bibinfo{journal}{Phys. Rev. D} \textbf{\bibinfo{volume}{15}},
  \bibinfo{pages}{2752} (\bibinfo{year}{1977}{\natexlab{a}}).

\bibitem[{\citenamefont{Gibbons and
  Hawking}(1977{\natexlab{b}})}]{Gibbons:1977mu}
\bibinfo{author}{\bibfnamefont{G.~W.} \bibnamefont{Gibbons}} \bibnamefont{and}
  \bibinfo{author}{\bibfnamefont{S.~W.} \bibnamefont{Hawking}},
  \bibinfo{journal}{Phys. Rev. D} \textbf{\bibinfo{volume}{15}},
  \bibinfo{pages}{2738} (\bibinfo{year}{1977}{\natexlab{b}}).

\bibitem[{\citenamefont{Hawking}(1975)}]{Hawking:1975vcx}
\bibinfo{author}{\bibfnamefont{S.~W.} \bibnamefont{Hawking}},
  \bibinfo{journal}{Commun. Math. Phys.} \textbf{\bibinfo{volume}{43}},
  \bibinfo{pages}{199} (\bibinfo{year}{1975}), \bibinfo{note}{[Erratum:
  Commun.Math.Phys. 46, 206 (1976)]}.

\bibitem[{\citenamefont{York}(1972)}]{York:1972sj}
\bibinfo{author}{\bibfnamefont{J.~W.} \bibnamefont{York}, \bibfnamefont{Jr.}},
  \bibinfo{journal}{Phys. Rev. Lett.} \textbf{\bibinfo{volume}{28}},
  \bibinfo{pages}{1082} (\bibinfo{year}{1972}).

\bibitem[{\citenamefont{Brown and York}(1994{\natexlab{a}})}]{Brown:1994su}
\bibinfo{author}{\bibfnamefont{J.~D.} \bibnamefont{Brown}} \bibnamefont{and}
  \bibinfo{author}{\bibfnamefont{J.~W.} \bibnamefont{York},
  \bibfnamefont{Jr.}}, in \emph{\bibinfo{booktitle}{{The Black Hole 25 Years
  After}}} (\bibinfo{year}{1994}{\natexlab{a}}), pp. \bibinfo{pages}{1--24},
  \eprint{gr-qc/9405024}.

\bibitem[{\citenamefont{Brown and York}(1993{\natexlab{a}})}]{Brown:1992bq}
\bibinfo{author}{\bibfnamefont{J.~D.} \bibnamefont{Brown}} \bibnamefont{and}
  \bibinfo{author}{\bibfnamefont{J.~W.} \bibnamefont{York},
  \bibfnamefont{Jr.}}, \bibinfo{journal}{Phys. Rev. D}
  \textbf{\bibinfo{volume}{47}}, \bibinfo{pages}{1420}
  (\bibinfo{year}{1993}{\natexlab{a}}), \eprint{gr-qc/9209014}.

\bibitem[{\citenamefont{Hawking and Page}(1983)}]{Hawking:1982dh}
\bibinfo{author}{\bibfnamefont{S.~W.} \bibnamefont{Hawking}} \bibnamefont{and}
  \bibinfo{author}{\bibfnamefont{D.~N.} \bibnamefont{Page}},
  \bibinfo{journal}{Commun. Math. Phys.} \textbf{\bibinfo{volume}{87}},
  \bibinfo{pages}{577} (\bibinfo{year}{1983}).

\bibitem[{\citenamefont{Martinez}(1995)}]{Martinez:1994ub}
\bibinfo{author}{\bibfnamefont{E.~A.} \bibnamefont{Martinez}},
  \bibinfo{journal}{Phys. Rev. D} \textbf{\bibinfo{volume}{51}},
  \bibinfo{pages}{5732} (\bibinfo{year}{1995}), \eprint{gr-qc/9412051}.

\bibitem[{\citenamefont{Bousso and Hawking}(1996)}]{Bousso:1996au}
\bibinfo{author}{\bibfnamefont{R.}~\bibnamefont{Bousso}} \bibnamefont{and}
  \bibinfo{author}{\bibfnamefont{S.~W.} \bibnamefont{Hawking}},
  \bibinfo{journal}{Phys. Rev. D} \textbf{\bibinfo{volume}{54}},
  \bibinfo{pages}{6312} (\bibinfo{year}{1996}), \eprint{gr-qc/9606052}.

\bibitem[{\citenamefont{Teitelboim}(2001)}]{Teitelboim:2001skl}
\bibinfo{author}{\bibfnamefont{C.}~\bibnamefont{Teitelboim}}, in
  \emph{\bibinfo{booktitle}{{Meeting on Strings and Gravity: Tying the Forces
  Together}}} (\bibinfo{year}{2001}), pp. \bibinfo{pages}{291--299},
  \eprint{hep-th/0203258}.

\bibitem[{\citenamefont{Poisson}(2009)}]{Poisson:2009pwt}
\bibinfo{author}{\bibfnamefont{E.}~\bibnamefont{Poisson}},
  \emph{\bibinfo{title}{{A Relativist's Toolkit: The Mathematics of Black-Hole
  Mechanics}}} (\bibinfo{publisher}{Cambridge University Press},
  \bibinfo{year}{2009}).

\bibitem[{\citenamefont{Hawking and Israel}(1979)}]{Hawking:1979ig}
\bibinfo{author}{\bibfnamefont{S.~W.} \bibnamefont{Hawking}} \bibnamefont{and}
  \bibinfo{author}{\bibfnamefont{W.}~\bibnamefont{Israel}},
  \emph{\bibinfo{title}{{General Relativity}: {An Einstein Centenary Survey}}}
  (\bibinfo{publisher}{Univ. Pr.}, \bibinfo{address}{Cambridge, UK},
  \bibinfo{year}{1979}), ISBN \bibinfo{isbn}{978-0-521-29928-2}.

\bibitem[{\citenamefont{Banks}(2001)}]{Banks:2000fe}
\bibinfo{author}{\bibfnamefont{T.}~\bibnamefont{Banks}}, \bibinfo{journal}{Int.
  J. Mod. Phys. A} \textbf{\bibinfo{volume}{16}}, \bibinfo{pages}{910}
  (\bibinfo{year}{2001}), \eprint{hep-th/0007146}.

\bibitem[{\citenamefont{Fischler}(2000)}]{fischler2000taking}
\bibinfo{author}{\bibfnamefont{W.}~\bibnamefont{Fischler}},
  \bibinfo{journal}{Talk given at role of scaling laws in physics and biology
  (Celebrating the 60th birthday of Geoffrey West), Santa Fe}
  \textbf{\bibinfo{volume}{19}} (\bibinfo{year}{2000}).

\bibitem[{\citenamefont{Banihashemi and Jacobson}(2022)}]{Banihashemi:2022jys}
\bibinfo{author}{\bibfnamefont{B.}~\bibnamefont{Banihashemi}} \bibnamefont{and}
  \bibinfo{author}{\bibfnamefont{T.}~\bibnamefont{Jacobson}},
  \bibinfo{journal}{JHEP} \textbf{\bibinfo{volume}{07}}, \bibinfo{pages}{042}
  (\bibinfo{year}{2022}), \eprint{2204.05324}.

\bibitem[{\citenamefont{Marolf}(2022)}]{Marolf:2022ybi}
\bibinfo{author}{\bibfnamefont{D.}~\bibnamefont{Marolf}},
  \bibinfo{journal}{JHEP} \textbf{\bibinfo{volume}{07}}, \bibinfo{pages}{108}
  (\bibinfo{year}{2022}), \eprint{2203.07421}.

\bibitem[{\citenamefont{Chandrasekaran
  et~al.}(2022)\citenamefont{Chandrasekaran, Penington, and
  Witten}}]{Chandrasekaran:2022eqq}
\bibinfo{author}{\bibfnamefont{V.}~\bibnamefont{Chandrasekaran}},
  \bibinfo{author}{\bibfnamefont{G.}~\bibnamefont{Penington}},
  \bibnamefont{and} \bibinfo{author}{\bibfnamefont{E.}~\bibnamefont{Witten}}
  (\bibinfo{year}{2022}), \eprint{2209.10454}.

\bibitem[{\citenamefont{Chandrasekaran
  et~al.}(2023)\citenamefont{Chandrasekaran, Longo, Penington, and
  Witten}}]{Chandrasekaran:2022cip}
\bibinfo{author}{\bibfnamefont{V.}~\bibnamefont{Chandrasekaran}},
  \bibinfo{author}{\bibfnamefont{R.}~\bibnamefont{Longo}},
  \bibinfo{author}{\bibfnamefont{G.}~\bibnamefont{Penington}},
  \bibnamefont{and} \bibinfo{author}{\bibfnamefont{E.}~\bibnamefont{Witten}},
  \bibinfo{journal}{JHEP} \textbf{\bibinfo{volume}{02}}, \bibinfo{pages}{082}
  (\bibinfo{year}{2023}), \eprint{2206.10780}.

\bibitem[{\citenamefont{Banados et~al.}(1994)\citenamefont{Banados, Teitelboim,
  and Zanelli}}]{Banados:1993qp}
\bibinfo{author}{\bibfnamefont{M.}~\bibnamefont{Banados}},
  \bibinfo{author}{\bibfnamefont{C.}~\bibnamefont{Teitelboim}},
  \bibnamefont{and} \bibinfo{author}{\bibfnamefont{J.}~\bibnamefont{Zanelli}},
  \bibinfo{journal}{Phys. Rev. Lett.} \textbf{\bibinfo{volume}{72}},
  \bibinfo{pages}{957} (\bibinfo{year}{1994}), \eprint{gr-qc/9309026}.

\bibitem[{\citenamefont{Gregory et~al.}(2014)\citenamefont{Gregory, Moss, and
  Withers}}]{Gregory:2013hja}
\bibinfo{author}{\bibfnamefont{R.}~\bibnamefont{Gregory}},
  \bibinfo{author}{\bibfnamefont{I.~G.} \bibnamefont{Moss}}, \bibnamefont{and}
  \bibinfo{author}{\bibfnamefont{B.}~\bibnamefont{Withers}},
  \bibinfo{journal}{JHEP} \textbf{\bibinfo{volume}{03}}, \bibinfo{pages}{081}
  (\bibinfo{year}{2014}), \eprint{1401.0017}.

\bibitem[{\citenamefont{Spradlin et~al.}(2001)\citenamefont{Spradlin,
  Strominger, and Volovich}}]{Spradlin:2001pw}
\bibinfo{author}{\bibfnamefont{M.}~\bibnamefont{Spradlin}},
  \bibinfo{author}{\bibfnamefont{A.}~\bibnamefont{Strominger}},
  \bibnamefont{and} \bibinfo{author}{\bibfnamefont{A.}~\bibnamefont{Volovich}},
  in \emph{\bibinfo{booktitle}{{Les Houches Summer School: Session 76: Euro
  Summer School on Unity of Fundamental Physics: Gravity, Gauge Theory and
  Strings}}} (\bibinfo{year}{2001}), pp. \bibinfo{pages}{423--453},
  \eprint{hep-th/0110007}.

\bibitem[{\citenamefont{Bardeen et~al.}(1973)\citenamefont{Bardeen, Carter, and
  Hawking}}]{Bardeen:1973gs}
\bibinfo{author}{\bibfnamefont{J.~M.} \bibnamefont{Bardeen}},
  \bibinfo{author}{\bibfnamefont{B.}~\bibnamefont{Carter}}, \bibnamefont{and}
  \bibinfo{author}{\bibfnamefont{S.~W.} \bibnamefont{Hawking}},
  \bibinfo{journal}{Commun. Math. Phys.} \textbf{\bibinfo{volume}{31}},
  \bibinfo{pages}{161} (\bibinfo{year}{1973}).

\bibitem[{\citenamefont{Banihashemi et~al.}(2022)\citenamefont{Banihashemi,
  Jacobson, Svesko, and Visser}}]{Banihashemi:2022htw}
\bibinfo{author}{\bibfnamefont{B.}~\bibnamefont{Banihashemi}},
  \bibinfo{author}{\bibfnamefont{T.}~\bibnamefont{Jacobson}},
  \bibinfo{author}{\bibfnamefont{A.}~\bibnamefont{Svesko}}, \bibnamefont{and}
  \bibinfo{author}{\bibfnamefont{M.}~\bibnamefont{Visser}}
  (\bibinfo{year}{2022}), \eprint{2208.11706}.

\bibitem[{\citenamefont{Blau et~al.}(1987)\citenamefont{Blau, Guendelman, and
  Guth}}]{Blau:1986cw}
\bibinfo{author}{\bibfnamefont{S.~K.} \bibnamefont{Blau}},
  \bibinfo{author}{\bibfnamefont{E.~I.} \bibnamefont{Guendelman}},
  \bibnamefont{and} \bibinfo{author}{\bibfnamefont{A.~H.} \bibnamefont{Guth}},
  \bibinfo{journal}{Phys. Rev. D} \textbf{\bibinfo{volume}{35}},
  \bibinfo{pages}{1747} (\bibinfo{year}{1987}).

\bibitem[{\citenamefont{Brown and York}(1993{\natexlab{b}})}]{Brown:1992br}
\bibinfo{author}{\bibfnamefont{J.~D.} \bibnamefont{Brown}} \bibnamefont{and}
  \bibinfo{author}{\bibfnamefont{J.~W.} \bibnamefont{York},
  \bibfnamefont{Jr.}}, \bibinfo{journal}{Phys. Rev. D}
  \textbf{\bibinfo{volume}{47}}, \bibinfo{pages}{1407}
  (\bibinfo{year}{1993}{\natexlab{b}}), \eprint{gr-qc/9209012}.

\bibitem[{\citenamefont{Brown and York}(1994{\natexlab{b}})}]{Brown:1993ke}
\bibinfo{author}{\bibfnamefont{J.~D.} \bibnamefont{Brown}} \bibnamefont{and}
  \bibinfo{author}{\bibfnamefont{J.~W.} \bibnamefont{York},
  \bibfnamefont{Jr.}}, \bibinfo{journal}{Math. Phys. Stud.}
  \textbf{\bibinfo{volume}{15}}, \bibinfo{pages}{23}
  (\bibinfo{year}{1994}{\natexlab{b}}), \eprint{gr-qc/9303012}.

\bibitem[{\citenamefont{Linde}(1984)}]{Linde:1983mx}
\bibinfo{author}{\bibfnamefont{A.~D.} \bibnamefont{Linde}},
  \bibinfo{journal}{Lett. Nuovo Cim.} \textbf{\bibinfo{volume}{39}},
  \bibinfo{pages}{401} (\bibinfo{year}{1984}).

\bibitem[{\citenamefont{Vilenkin}(1984)}]{Vilenkin:1984wp}
\bibinfo{author}{\bibfnamefont{A.}~\bibnamefont{Vilenkin}},
  \bibinfo{journal}{Phys. Rev. D} \textbf{\bibinfo{volume}{30}},
  \bibinfo{pages}{509} (\bibinfo{year}{1984}).

\bibitem[{\citenamefont{Vilenkin}(1982)}]{Vilenkin:1982de}
\bibinfo{author}{\bibfnamefont{A.}~\bibnamefont{Vilenkin}},
  \bibinfo{journal}{Phys. Lett. B} \textbf{\bibinfo{volume}{117}},
  \bibinfo{pages}{25} (\bibinfo{year}{1982}).

\bibitem[{\citenamefont{Hartle and Hawking}(1983)}]{Hartle:1983ai}
\bibinfo{author}{\bibfnamefont{J.~B.} \bibnamefont{Hartle}} \bibnamefont{and}
  \bibinfo{author}{\bibfnamefont{S.~W.} \bibnamefont{Hawking}},
  \bibinfo{journal}{Phys. Rev. D} \textbf{\bibinfo{volume}{28}},
  \bibinfo{pages}{2960} (\bibinfo{year}{1983}).

\bibitem[{\citenamefont{Brown and Dahlen}(2012)}]{Brown:2011gt}
\bibinfo{author}{\bibfnamefont{A.~R.} \bibnamefont{Brown}} \bibnamefont{and}
  \bibinfo{author}{\bibfnamefont{A.}~\bibnamefont{Dahlen}},
  \bibinfo{journal}{Phys. Rev. D} \textbf{\bibinfo{volume}{85}},
  \bibinfo{pages}{104026} (\bibinfo{year}{2012}), \eprint{1111.0301}.

\bibitem[{\citenamefont{Witten}(1982)}]{Witten:1981gj}
\bibinfo{author}{\bibfnamefont{E.}~\bibnamefont{Witten}},
  \bibinfo{journal}{Nucl. Phys. B} \textbf{\bibinfo{volume}{195}},
  \bibinfo{pages}{481} (\bibinfo{year}{1982}).

\bibitem[{\citenamefont{Maldacena}(2010)}]{Maldacena:2010un}
\bibinfo{author}{\bibfnamefont{J.}~\bibnamefont{Maldacena}}
  (\bibinfo{year}{2010}), \eprint{1012.0274}.

\bibitem[{\citenamefont{Callan and Coleman}(1977)}]{Callan:1977pt}
\bibinfo{author}{\bibfnamefont{C.~G.} \bibnamefont{Callan}, \bibfnamefont{Jr.}}
  \bibnamefont{and} \bibinfo{author}{\bibfnamefont{S.~R.}
  \bibnamefont{Coleman}}, \bibinfo{journal}{Phys. Rev. D}
  \textbf{\bibinfo{volume}{16}}, \bibinfo{pages}{1762} (\bibinfo{year}{1977}).

\bibitem[{\citenamefont{Cespedes et~al.}(2021)\citenamefont{Cespedes, de~Alwis,
  Muia, and Quevedo}}]{Cespedes:2020xpn}
\bibinfo{author}{\bibfnamefont{S.}~\bibnamefont{Cespedes}},
  \bibinfo{author}{\bibfnamefont{S.~P.} \bibnamefont{de~Alwis}},
  \bibinfo{author}{\bibfnamefont{F.}~\bibnamefont{Muia}}, \bibnamefont{and}
  \bibinfo{author}{\bibfnamefont{F.}~\bibnamefont{Quevedo}},
  \bibinfo{journal}{Phys. Rev. D} \textbf{\bibinfo{volume}{104}},
  \bibinfo{pages}{026013} (\bibinfo{year}{2021}), \eprint{2011.13936}.

\bibitem[{\citenamefont{Bachlechner}(2016)}]{Bachlechner:2016mtp}
\bibinfo{author}{\bibfnamefont{T.~C.} \bibnamefont{Bachlechner}},
  \bibinfo{journal}{JHEP} \textbf{\bibinfo{volume}{12}}, \bibinfo{pages}{155}
  (\bibinfo{year}{2016}), \eprint{1608.07576}.

\bibitem[{\citenamefont{Susskind}(2021)}]{Susskind:2021yvs}
\bibinfo{author}{\bibfnamefont{L.}~\bibnamefont{Susskind}}
  (\bibinfo{year}{2021}), \eprint{2107.11688}.

\bibitem[{\citenamefont{De~Alwis et~al.}(2020)\citenamefont{De~Alwis, Muia,
  Pasquarella, and Quevedo}}]{DeAlwis:2019rxg}
\bibinfo{author}{\bibfnamefont{S.~P.} \bibnamefont{De~Alwis}},
  \bibinfo{author}{\bibfnamefont{F.}~\bibnamefont{Muia}},
  \bibinfo{author}{\bibfnamefont{V.}~\bibnamefont{Pasquarella}},
  \bibnamefont{and} \bibinfo{author}{\bibfnamefont{F.}~\bibnamefont{Quevedo}},
  \bibinfo{journal}{Fortsch. Phys.} \textbf{\bibinfo{volume}{68}},
  \bibinfo{pages}{2000069} (\bibinfo{year}{2020}), \eprint{1909.01975}.

\bibitem[{\citenamefont{Farhi et~al.}(1990)\citenamefont{Farhi, Guth, and
  Guven}}]{Farhi:1989yr}
\bibinfo{author}{\bibfnamefont{E.}~\bibnamefont{Farhi}},
  \bibinfo{author}{\bibfnamefont{A.~H.} \bibnamefont{Guth}}, \bibnamefont{and}
  \bibinfo{author}{\bibfnamefont{J.}~\bibnamefont{Guven}},
  \bibinfo{journal}{Nucl. Phys. B} \textbf{\bibinfo{volume}{339}},
  \bibinfo{pages}{417} (\bibinfo{year}{1990}).

\bibitem[{\citenamefont{Coleman and De~Luccia}(1980)}]{Coleman:1980aw}
\bibinfo{author}{\bibfnamefont{S.~R.} \bibnamefont{Coleman}} \bibnamefont{and}
  \bibinfo{author}{\bibfnamefont{F.}~\bibnamefont{De~Luccia}},
  \bibinfo{journal}{Phys. Rev. D} \textbf{\bibinfo{volume}{21}},
  \bibinfo{pages}{3305} (\bibinfo{year}{1980}).

\bibitem[{\citenamefont{Fischler et~al.}(1990)\citenamefont{Fischler, Morgan,
  and Polchinski}}]{Fischler:1990pk}
\bibinfo{author}{\bibfnamefont{W.}~\bibnamefont{Fischler}},
  \bibinfo{author}{\bibfnamefont{D.}~\bibnamefont{Morgan}}, \bibnamefont{and}
  \bibinfo{author}{\bibfnamefont{J.}~\bibnamefont{Polchinski}},
  \bibinfo{journal}{Phys. Rev. D} \textbf{\bibinfo{volume}{42}},
  \bibinfo{pages}{4042} (\bibinfo{year}{1990}).

\end{thebibliography}

\end{document}